\documentclass{aa}  
\usepackage{graphicx}
\usepackage{natbib}
\usepackage[dvipsnames]{xcolor}
\usepackage{placeins}

\usepackage{txfonts}
\usepackage[colorlinks=true, allcolors = blue]{hyperref}
\begin{document} 

   \title{2D unified atmosphere and wind simulations \\ for a grid of O-type stars}


   \author{N. Moens \inst{1}
          \and
          D. Debnath \inst{1}
          \and
          O. Verhamme \inst{1}
          \and
          F. Backs \inst{1}
          \and
          C. Van der Sijpt \inst{1}
          \and
          J.O. Sundqvist \inst{1}
          \and
          A.A.C. Sander\inst{\ref{inst:ari},\ref{inst:iwr}}
          }

    \institute{Instituut voor Sterrenkunde, KU Leuven,
              Celestijnenlaan 200D, 3001 Leuven, Belgium,\\
              \email{nicolas.moens@kuleuven.be}
            \and
            {Zentrum f{\"u}r Astronomie der Universit{\"a}t Heidelberg, Astronomisches Rechen-Institut, M{\"o}nchhofstr. 12-14, 69120 Heidelberg, \\Germany\label{inst:ari}}
            \and
    {Universit\"at Heidelberg, Interdiszipli\"ares Zentrum f\"ur Wissenschaftliches Rechnen, 69120 Heidelberg, Germany\label{inst:iwr}}
            }

   \date{Received xxxx xx, xxxx; accepted xxxx xx, xxxx}

 
  \abstract
   {
   The atmospheres of massive O-type stars (O stars) are dynamic, turbulent environments resulting from radiatively driven instabilities over the iron bump, located slightly beneath the stellar surface. Here, complex radiation hydrodynamic processes affect the structure of the atmosphere as well as the formation of spectral lines. In quantitative spectroscopic analysis, the effects of these processes are often parametrized with ad hoc techniques and values.
   }
   {This work is aimed at exploring how variation of basic atmospheric parameters affects
   the dynamics within the subsurface turbulent zone. We also explore how this turbulence relates to absorption lines formed in the photosphere for a broad range of O stars at solar metallically.
   }
   {
   
   The work in this paper centers around a grid of 2D, radiation-hydrodynamic O-star atmosphere and wind simulations, where the turbulent region is an emergent property of the simulation. For each of the 36 models in the grid, we derived the turbulent properties and correlated them to an estimate of turbulent line broadening imposed by the models' velocity fields.
   }
   {
   Our work 
   suggests that the subphotospheric 
   turbulent velocity in O-stars scales approximately with the square of the Eddington parameter, $\Gamma_{\rm e}$. We also find
   a linear correlation between subphotospheric turbulent velocity and the line broadening of several synthetic photospheric absorption lines. Radiation carries more energy than advection throughout the atmosphere for all models in the grid; however, for O-type supergiants, the latter can account for up to 30 \% of the total flux at the peak of the iron bump.
   }
   {}


   \maketitle
%

\section{Introduction}

The spectroscopic analysis of O-type stars (hereafter, O stars) is significantly influenced by the presence of strong, line-driven wind outflows \citep[e.g.,][]{Lucy1970,CAK1975}. Any accurate spectroscopic modeling of these stars thus requires a simultaneous analysis of both the deeper atmospheric layers and the stellar wind \citep[e.g.,][]{Gabler1989}.
Standard assumptions for such expanding atmosphere models are stationarity and spherical symmetry \citep[e.g.,][]{Santolaya-Rey1997, grafener2002, Hillier1998}.
Over the years, these models have typically begun to employ increasingly complex parameterizations for structure formation, primarily focusing on the stellar wind part of the atmosphere \citep{Eversberg1998, Puls2006, Oskinova2007, Zsargo2008, Sundqvist2018}, driven by observational needs (e.g., the \ion{P}{v} problem \citealt{Oskinova2007, Fullerton2006, Surlan2013, Sundqvist2010}) 
as well as theoretical requirements 
\citep{Feldmeier1995}.

In addition to wind variability, observations of photospheric absorption lines have indicated broadening processes due to turbulent motions beneath the photosphere \citep{Conti1977, Howarth1997}, acting in addition to broadening due to stellar rotation \citep{Markova2014}. Theoretical predictions \citep{Blaes2003,  Cantiello2009, VanderSijpt2025} and numerical simulations \citep{Jiang2015, Debnath2024} also suggest complex dynamics deeper within the typically assumed radiative outer envelope of massive stars. 
These dynamics, supported by instabilities over the iron opacity bump \citep{Stothers1993,Iglesias1996}, have been shown to alter the stellar structure in the outer envelope and atmosphere \citep{Debnath2024,Gonzalez-Tora2025,Gonzalez-Tora2025a}. 
This dynamic region just below the photosphere has been proposed as one possible mechanism behind the macro-turbulent broadening of observed spectral lines \citep{Cantiello2021}. Other 
suggested origins include heat-driven processes \citep{Aerts2009}, stochastic gravity waves \citep{Aerts2015}, or a combination of these factors \citep{Bowman2020}.

While the exact structure formation mechanism is uncertain, several instabilities in radiation-pressure-dominated environments have been researched in the literature. \citet{Shaviv2001} considered global instabilities in Thomson scattering atmospheres close to the Eddington limit, while \citet{Blaes2003} examined local, optically thick radiation-hydrodynamical (RHD) instabilities, and \citet{Jacquet2011} and \citet{Jiang2013} investigated Rayleigh-Taylor instabilities in radiation-dominated gases.
Additionally, strong opacity gradients could lead to so-called strange modes \citep{Gautschy1990, Glatzel1994, Saio1998}. 
Recently, these aforementioned mechanisms have been evaluated by \citet{VanderSijpt2025} for the same type of simulations as described in this work and by \citet{Moens2022} and \citet{Debnath2024}. 
The result of this study showed that the simulated structure growth does not follow predictions for the gravity-mode instability; however, the results were  inconclusive with respect to the importance of strange modes 
in maintaining a structured atmosphere, 
hinting at Rayleigh-Taylor-like instabilities potentially being responsible for the breakup of initial conditions. 

In this study, we utilized multidimensional (multi-D) unified atmosphere and wind models for O stars within \texttt{mpi-amrvac} \citep{Keppens2023}, employing the flux-limited diffusion (FLD) radiation-hydrodynamics (RHD) module \citep{Moens2021}. 
The multi-D models used within this framework \citep{Moens2022, Debnath2024} self-consistently generate turbulent regions below the photosphere due to atmospheric instabilities \citep{VanderSijpt2025}.

A preliminary theoretical analysis on spectral synthesis of turbulent multi-D O star atmosphere simulations has been conducted by \citet{Schultz2022, Schultz2023} using Monte Carlo techniques on variable-Eddington tensor method models by \citet{Jiang} 
, and very recently also by \citet{Delbroek2025} using extensions of the aforementioned unified atmosphere and wind simulations by \citet{Moens2022} and \citet{Debnath2024}. However, these studies were limited to a small number of stellar models. With our computationally efficient FLD model, we aim to characterize the dynamical atmospheric properties across a broad range of massive-star models. In this paper, we map the turbulent properties of the outermost layers of Galactic O stars. Additionally, we seek to link these properties to the broadening of photospheric absorption induced by the atmospheric velocity fields.
In the future, findings from our multi-D simulations could also be
integrated into 1D stellar atmosphere models (e.g., \citealt{Puls2005, Hillier1991, Sander2015}) and evolution codes (e.g., \citealt{Jermyn2023}). An approximated integration has already been initiated for a few models in the stellar atmosphere code PoWR \citep{Gonzalez-Tora2025a, Gonzalez-Tora2025} with further improvements making up a work in progress.

\section{Grid of 2D O star models} \label{sec: methods}

This work is focused on analyzing a grid of 2D models for the atmosphere and winds of galactic O stars. In the following, we describe the basics of the individual models as well as the stellar parameter range of the grid.

\subsection{2D atmosphere and wind model setup} \label{ssec: Setup}
The individual O star models are based on the general RHD-FLD \texttt{mpi-amrvac} method developed by \citet{Moens2021}, initially used to model the outflows of WR stars \citep{Moens2022} and later  adapted for the deep turbulent atmospheres and winds of O stars \citep{Debnath2024}.
We solved the hydrodynamic (HD) equations along with the zeroth-moment equation of the radiative transfer equation (RTE)
\begin{eqnarray}
    \partial_t \rho + \nabla \cdot (\rho \vec{\varv}) &=& 0 \label{eq: hd_rho}, \\
    \partial_t (\vec{\varv} \rho) + \nabla \cdot (\vec{\varv} \rho \vec{\varv} + p\vec{I}) &=& - \vec{f_g} + \vec{f_r} \label{eq: hd_mom}, \\
    \partial_t e + \nabla \cdot (e \vec{\varv} + p \vec{\varv}) &=& - \vec{f_g} \cdot \vec{\varv} + \vec{f_r} \cdot \vec{\varv} + \dot{q} \label{eq: hd_e} , \\
    \partial_t E + \nabla \cdot (E \vec{\varv} + F) &=& \nabla \vec{\varv} : P - \dot{q} \label{eq: hd_erad}.
\end{eqnarray}

Here, $\rho$ is the gas density, $\vec{\varv}$ is the gas velocity, $p\vec{I}$ is the scalar gas pressure times the identity matrix, $e$ is the total gas energy density, and $E$ is the radiation energy density. 
The source terms in the momentum equation include a point-source-like gravitational force, $\vec{f}_{\rm g}$, and the radiation force, $\vec{f}_{\rm r}$. 
The source terms in the gas energy equation consist of the work done by the gravitational and radiation forces and a heating and cooling term, $\dot{q}$.
The equations are solved in a Cartesian box-in-star
setup, with spherical flux correction terms according to Appendix A in \citet{Moens2021}.
The gas system is closed with the ideal gas law 
and the radiation system with the FLD closure relations for radiation flux 
and radiation pressure 
as described in \citet{Moens2021}.
As we consider the outermost layers of the star, the atmosphere's self-gravity can be safely ignored. 
Thus, the gravitational term in the momentum and gas energy equations can be approximated by point-mass gravity with a stellar mass, $M_\ast$.
The radiation force is calculated with the FLD flux as 
\begin{equation} 
\vec{f_r} = \rho \frac{\kappa_F\vec{F}}{c}, 
\end{equation} 
and the radiative heating and cooling terms are expressed as 
\begin{equation}
\dot{q} = c \kappa_E \rho E - 4 \pi \kappa_B \rho B(T_g),
\end{equation}
where $\kappa_F$, $\kappa_E$, and $\kappa_B$ are the frequency-integrated flux mean, energy mean, and Planck mean opacities. While the work is in progress to compute these different weighted opacities, we currently considered the same value for all three. 
We used a "hybrid" opacity scheme, where the total opacity is computed as the sum of tabulated Rosseland mean opacities and Sobolev-enhanced line-driving opacities calculated from a large atomic database (see also  \citealt{Poniatowski2022, Castor2004} for more details). The latter represents the effect of Doppler-shift enhanced spectral lines which is not represented in the assumptions of a Rosseland mean. As the Rosseland mean opacities dominate in optically thick regions and naturally revert to electron scattering opacities outside of the optically thick stellar photosphere, while the Sobolev enhanced opacities naturally revert to zero in optically thick regions, the transition between the two happens smoothly and parameter-free when using their sum and there should be minimal overlap (See also \citealt{Castor2004}, Ch. 6).
In the line force, the finite disk correction \citep{Pauldrach1986} was further used to account for geometric effects of the stellar disk. 
For an exact description of the implementation, we refer to \citet{Debnath2024}.

Integrating Eqs. \eqref{eq: hd_rho}-\eqref{eq: hd_erad} over time on a 2D finite volume grid leads to a potentially turbulent atmosphere with a self-consistent stellar wind driven by line-driving opacities. Here, the mass-loss rate is not set, but computed via self-consistent calculation instead. 
The atmosphere models are computed time-dependently on a grid ranging from $r=R_\text{c}$ to $r=3R_\text{c}$, where $R_\text{c}$ is the set lower boundary radius chosen to approximately coincide with a temperature of $500~\mathrm{kK}$. 
This is well below both the photosphere and the iron recombination bump seen in Rosseland mean opacities \citep{Iglesias1996},  thereby partially covering the stable, radiative zone in the stellar envelope. This has proven important for correctly treating boundary conditions, as the iron opacity bump is responsible for structure formation in the stellar atmosphere \citep{VanderSijpt2025}. Our deeper lower boundary conditions allow us to simulate the transition from a purely radiative region to the radii where structure formation starts.
Our models laterally extend $0.4 R_\text{c}$. 
By manual tuning, this has shown to be sufficient to cover multiple turbulent cells. With the help of a refined mesh, we resolved the dynamically interesting regions to $\sim 10^{-3} R_\text{c}$ in both the radial and lateral directions. This is sufficient to resolve the scale height of the rapidly decaying exponential density and pressure profiles around the photosphere.
The initial conditions are computed from a simplified spherically symmetric, steady-state 1D model (see \citealt{Debnath2024}). This model combines a hydrostatic atmosphere, stitched to a "$\beta$" velocity-law wind with given initial wind parameters.
The lateral boundary conditions are periodic.

\subsection{The grid} \label{ssec: Grid}

To model the atmospheric dynamics over a range of O stars quantitatively, a grid of models can be constructed by varying a set of input parameters: the stellar mass, $M_\ast$, classical Eddington ratio, $\Gamma_\text{e}$, and a lower boundary radius. Our model grid focuses on O stars on or near the main sequence, essentially following the grid calculations by \citet{Bjorklund2021}, which, in turn, were based on the spectroscopic O star calibration by \citet{Martins2005} using the 1D NLTE stellar atmosphere code CMFGEN \citep{Hillier1998}.
The grid contains input parameters for stars  in 1D non-convective radiative equilibrium models, corresponding to 12 spectroscopic dwarfs, 12 giants, and 12 supergiants. The ranges of Eddington parameters are $\Gamma_\text{e} = [0.07, 0.31]$ for the dwarfs, $\Gamma_\text{e} = [0.17, 0.38]$ for the giants, and $\Gamma_\text{e} = [0.27, 0.4]$ for the supergiants. 
All input parameters for the models are listed in Appendix \ref{app: Grid io} in order of rising mass, named D1 to D12 for the dwarfs, G1 to G12 for the giants, and S1 to S12 for the supergiants. 

\begin{figure} 
\centering 
\includegraphics[width=\hsize]{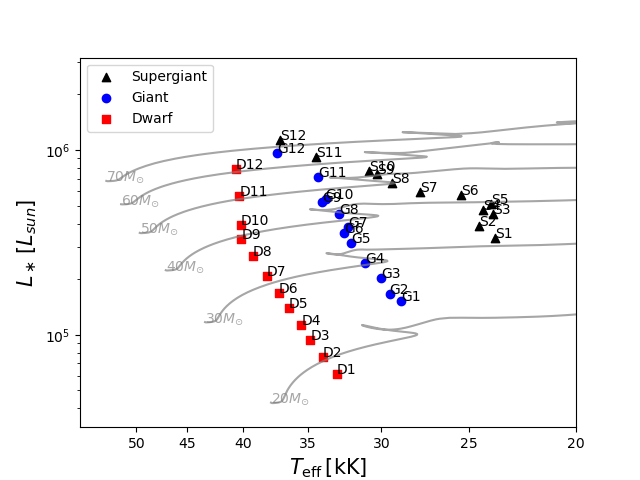} 
\caption{
HRD positions after model relaxation for the entire grid. The gray lines indicate MIST \citep{Dotter2016} evolution tracks for non-rotating massive stars. 
The effective temperatures shown here are emergent properties of the simulation. Due to statistical scatter in very turbulent models, they do not line up in perfect sequences.
\label{fig: HRD}}
\end{figure}

For this work, we considered only Galactic stars with solar abundances according to \citet{Grevesse1993}. 
One aspect of our modeling technique is that we cannot know in advance what effective temperature and photospheric radius the 2D atmosphere will end up relaxing into.
The initial conditions for our simulations are created using a simplified, purely radiative, 1D stellar atmosphere based on an assumed set of parameters that include the photospheric radius, stellar luminosity, and stellar mass. Integration over the hydrostatic equilibrium equation in the lower atmosphere then gives the hydrodynamic conditions at the lower boundary, which here is located at a chosen temperature of $500~\mathrm{kK}$. 
As the simulation progresses, the atmosphere evolves, excites turbulent layers, and, finally, stochastically relaxes to an atmosphere with a new photospheric radius and temperature. Consequently, the final model may have a different position on the HRD compared to the 1D model used as an initial condition. The new  corresponding, relaxed stellar parameters for each model can also be found in Appendix \ref{app: Grid io}.
After relaxing the simulations, their new average photospheric radii and effective temperatures were computed, and their position on the HRD was marked (see Fig. \ref{fig: HRD}). The exact procedure for calculating these parameter averages is explained in Sect. \ref{ssec: Stellar params} and the values are listed in Table \ref{table: output} of Appendix \ref{app: Grid io}.

For each model, the simulation must run long enough so that the initial conditions do not affect the general atmospheric dynamics. 
After the model relaxation, we need to sample a sufficiently broad range in time to calculate suitable spatial and temporal average quantities. In practice, this means we needed to continue the simulation for at least an additional 60 wind dynamical timescales; all time averages used in the rest of this paper are taken over these final 60 wind dynamical timescales. Appendix \ref{app: timescales} lists characteristic values of the different important timescales and how they are estimated in this work. 

\section{Dynamical properties} \label{sec: results}

Overall, in comparison to radiative 1D models, the turbulent motions in the unified stellar envelopes and winds simulated here can alter the averaged atmospheric structure properties primarily via two mechanisms: turbulent pressure support and enthalpy energy transport (convection). In this section, we aim to quantify these two mechanisms and how they affect the atmospheric structure throughout our grid.

\subsection{Relaxed atmospheric structure} \label{ssec: Average Struc}

\begin{figure*}
\centering
\includegraphics[width=0.49 \linewidth]{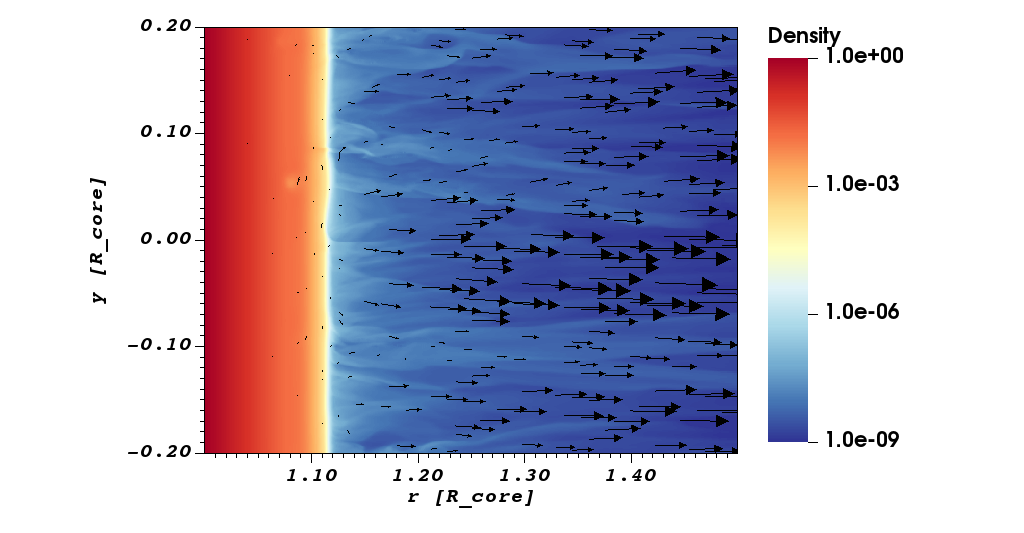}
\includegraphics[width=0.49 \linewidth]{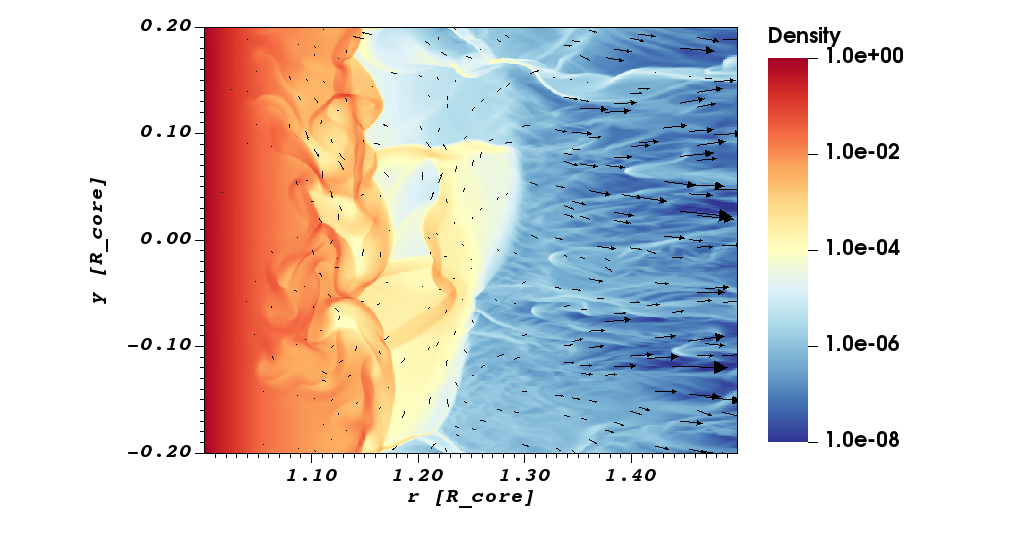}
  \caption{
  Density maps of a relatively steady atmosphere (left),  corresponding to a dwarf model D1  and a relatively turbulent atmosphere (right), corresponding to a supergiant model S12. The velocity field is indicated by the black arrows.
  The density is given in units of the lower boundary density (see Table \ref{table: input}), and the magnitude of the velocity peaks at $300 \, \rm km/s$
  }
     \label{fig: Density_fields}
\end{figure*}

Unlike 1D atmospheric models, the 2D RHD simulations discussed in this paper give rise to a  complex atmosphere and wind structure (see Fig. \ref{fig: Density_fields}). Across the grid, most models share a similar structure once relaxed: deep below the iron bump, for $T > 300~\mathrm{kK}$, the atmosphere is quasi-hydrostatic and energy transport is nearly purely radiative. Here, the density, energy, and radiation fields are 
smooth, as in a classical exponential atmosphere 
(see Fig. \ref{fig: Density_fields}). The velocity field is close to zero in all directions. 
Once the opacity starts increasing in the iron-opacity region, however, local parcels of gas get gravitationally unbound and shoot up toward the upper and cooler atmosphere. There, the opacity decreases again and the gas parcels start losing their upwards momentum. Depending on 
the local atmospheric conditions of these upper layers then, some parcels will experience significant line-driving there and re-accelerate, whereas others will turn over and sink toward the deeper atmosphere. 
We note that it is not the adiabatic cooling of the gas that causes the turnover, but the change in opacity due to a change in temperature. 
The atmosphere in the wind launching mechanism near the photosphere induces structures in the wind further out.

The large momentum of these gas cells takes the upper atmosphere out of 
hydrostatic equilibrium, and the gas becomes very dynamic and turbulent (see Fig. \ref{fig: Density_fields}).
Here, the turbulence is responsible for significant additional pressure in the atmosphere (see also Figs. 13 and 14 in \citealt{Debnath2024}). This turbulent pressure increases the effective scale height of the atmosphere.
Around the average photosphere, line-driving takes effect and launches an on average supersonic outflow. 
The wind is highly structured due to the turbulent and dynamically active launching region. 

\subsection{Effective temperature, photospheric radius and stellar luminosity} \label{ssec: Stellar params}

As mentioned in Sect. \ref{ssec: Grid}, in our simulations the stellar radii and effective temperatures are emergent and time-variable properties. 
Here, we define an average photospheric radius by first calculating, for each radial ray, the point where the Rosseland mean optical depth is two-thirds and then taking a lateral and time average
\begin{equation} 
R_{\rm ph} = \left< r(\tau=2/3) \right>. 
\end{equation}

The effective temperature is calculated from the average radiative flux at these $\tau=2/3$ points, 
\begin{equation} 
T_{\rm eff} = \left( \frac{ \left< F^{\rm obs}_r(\tau=2/3) \right> }{\sigma} \right)^{1/4}, 
\end{equation} 
where $\sigma$ is the Stefan-Boltzmann constant.\footnote{The observer's frame radiative flux is actually $\vec{F}^\text{obs} = \vec{F} + \vec{\varv}(E+P)$, where $\vec{F}$ is the corresponding co-moving frame flux. In the O star models presented here, however, $\vec{F} \gg \vec{\varv} E$ at the photosphere throughout the grid.}

The average model luminosity is then obtained from 
\begin{equation} 
L_\ast = 4 \pi \sigma R_{\rm ph}^2 T_{\rm eff}^4. 
\end{equation}

There are several methods to define an average stellar scaling radius in a spherical system. For example,  \citet{Schultz2023} compare the method above to one in which $T_{\rm eff}$ is defined from identifying, where the gas and flux temperatures are equal (which would be identical to the definition above in a 1D radiative model). They find that these two different effective temperature interpretations may significantly differ for massive stars close to the Eddington limit. In the O star grid presented here, different methods typically agree to within $0.1 R_\odot$ for the photospheric radius and well within $1~\mathrm{kK}$ for the effective temperature.

\subsection{Turbulent pressure} \label{ssec: Turbulent pressure}

We mainly focus on the effects of turbulent pressure in the atmospheric layers below the average wind sonic point, as in these layers there is no significant net outflow and turbulent velocities are approximately isotropic. Above this average sonic point, line-driven wind dynamics start to dominate and the turbulent motion can no longer assumed to be isotropic. Thus, for the region of interest, we can calculate an average turbulent pressure as a function of radius by averaging over time and lateral direction (see also \citealt{Jiang2023, Debnath2024})
\begin{equation} 
\overline{P_\text{turb}} = \left< \rho \varv_r^2 \right>, \label{eq: p_turb} 
\end{equation} 
such that the corresponding average turbulent velocity is 
\begin{equation} 
\overline{\varv_{\rm turb}} = \sqrt{\frac{\overline{P_\text{turb}}}{\left< \rho \right>}}. \label{eq: v_turb} 
\end{equation}

\begin{figure}
\centering
\includegraphics[width=\hsize]{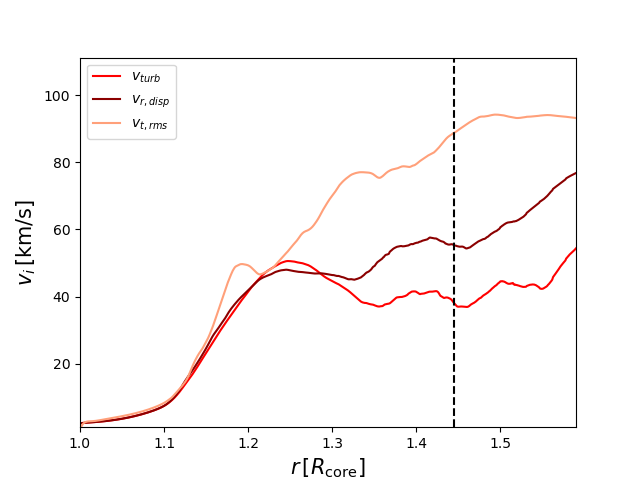}
  \caption{Value for the turbulent velocity as a function of radius for model G1. The vertical black line indicates the average model photosphere. The brown and pink lines indicate the radial and lateral velocity dispersions . The black dashed line represents the average photosphere.
  }
     \label{fig: vtrub_prof}
\end{figure}

This quantity characterizes typical velocities of the gas in the turbulent, approximately zero-net outflow region of the atmosphere, and can be 
calculated for each model in the grid. Figure \ref{fig: vtrub_prof} 
shows this quantity as a function of radius for a giant model (G1), with a mass of $21.1 \, M_\odot$, luminosity of $1.5 \cdot 10^5\,L_\odot$, and effective temperature of $29\,$kK. 

Evaluating the maximum value 
of the turbulent velocity below the photosphere\footnote{One could also consider the value at the photosphere to get a similar relation, but it appears that these two values are very strictly correlated.} over the entire grid shows a clear correlation with the classical Eddington parameter, $\Gamma_\text{e}$. 
For the dwarf and giant models, this relation is quite tight, with more spread in the supergiants.
This can be seen in Fig. \ref{fig: vturb_grid}, where the gray dotted line indicates the nearly quadratic, best-fit power law,
\begin{equation} 
\varv_{\rm turb} \sim \left(\Gamma_\text{e} \right)^{1.96}.
\end{equation}

For the lowest $\Gamma_\text{e}$ models, the calculated turbulent velocities seem to be dominated mainly by global, time-domain fluctuations in the atmospheric structure. These motions do not break up into turbulent cells and are perhaps related to so-called Lamb-oscillations induced by line-driving \citep{Key2025}.
As such, the $\varv_{\rm turb}$ values at the low $\Gamma_\text{e}$-end should be interpreted as an upper limit. 
We also note that the exact fit provided here might be subject to change when looking at more accurate 3D models (see discussion in Sect. \ref{sssec: 3D}). 

\begin{figure}
\centering
\includegraphics[width=\hsize]{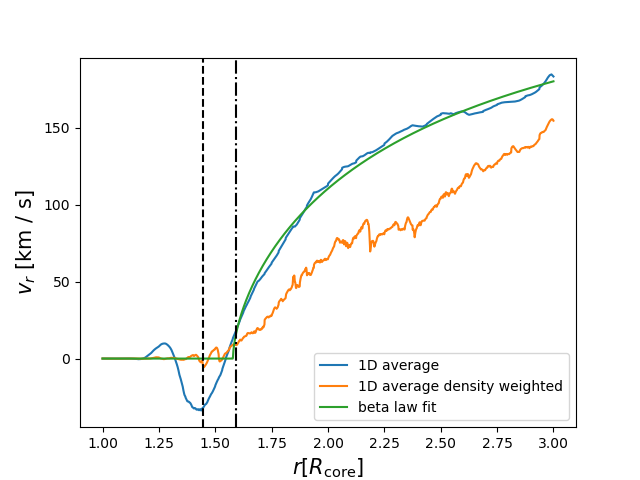}
  \caption{
   Average radial velocity profile as a function of radius (blue), the average density weighted radial velocity profile (orange), and the best fit $\beta$-velocity law (green) for the G1 model. The black dashed and dash-dotted lines represent the average photosphere and sonic radius, respectively.
  }
     \label{fig: Beta_vel}
\end{figure}

\begin{figure}
\centering
\includegraphics[width=\hsize]{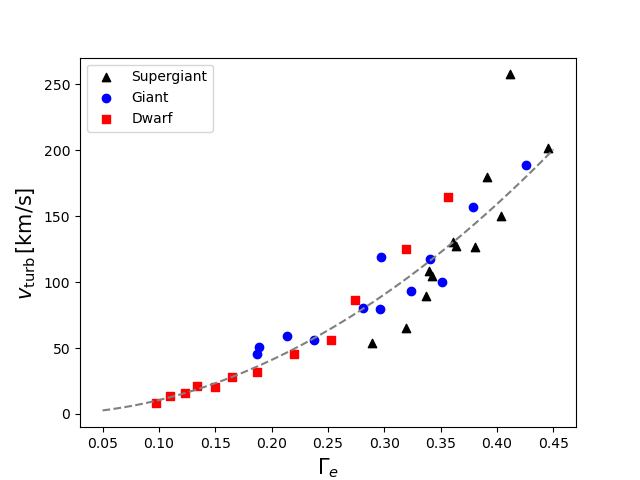}
  \caption{Maximum value of the turbulent velocity below the photosphere, as a function of the classical Eddington factor for the entire grid of models. Red squares indicate the series of dwarf models, blue circles the series of giants, and black triangles the series of supergiants.}
     \label{fig: vturb_grid}
\end{figure}

When comparing the turbulent perturbations in the velocity field, we find that the average magnitude of the velocity components is slightly bigger in the tangential direction as compared to the component in the radial direction. For most of the models, the ratio of the maximum tangential over radial velocity dispersion below the photosphere lies between one and two. 
This is less than what was found by \citet{Schultz2022}, but confirms the dominance by tangential motions. The reason for the relatively increased radial velocities as compared to \citet{Schultz2022} could be the included radially dominant line force. This small anisotropy is not a major problem for our definition of the turbulent velocity in Eq. \eqref{eq: p_turb}, since the expression is valid for a spherical configuration with net-zero tangential velocity and where the total pressure scale height is much smaller than the radius (see \citealt{Jiang2023} and \citealt{Debnath2024}).

\subsection{Convective energy transport} \label{ssec: Convective transport}

\begin{figure} 
\centering 
\includegraphics[width=\hsize]{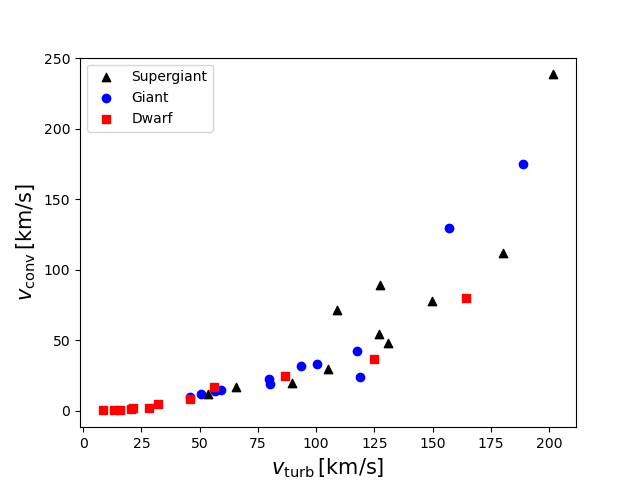} 
\caption{Correlation between the maximum convective velocity below the photosphere and the maximum turbulent velocity below the photosphere. Red squares indicate the series of dwarf models, blue circles the series of giants, and black triangles the series of supergiants. 
}
\label{fig: vturb_vconv} 
\end{figure}

By combining the gas and radiation energy equations, we can formulate an expression for the total energy flux, 
\begin{equation} 
\partial_t \left(e + E\right) + \nabla \cdot \vec{F}_{\rm tot} = 0, 
\end{equation} 
where $\vec{F}_{\rm tot}$ consists of the co-moving frame (diffusive) radiative flux, the radiative- and gas-enthalpic fluxes, the kinetic energy flux, and the gravito-potential flux. On average, we expect this total flux to be zero in the lateral directions. In the radial direction, however, we have 
\begin{equation} 
F_{\rm tot} = \varv_r \left( p + P + e + E - \rho \frac{G M_*}{r} \right) + F, 
\end{equation} 
where $F$ is the diffusive co-moving frame flux, here evaluated from the analytic FLD closure relation. The total flux is related to the total luminosity of the model $L_{\rm tot} = 4 \pi r^2 F_{\rm tot}$. At the lower boundary where the radial velocity is close to zero, it is dominated by the radiative luminosity.
From this expression, we compute the average convective energy flux carried by advection, 
\begin{equation} 
\overline{F_{\rm conv}} = \left< \varv_r \left( p + P + e + E \right) \right>. \label{eq: F_conv} 
\end{equation} 
Accordingly, the typical velocity associated with convective (enthalpy) energy transport (see also \citealt{Debnath2024})
\begin{equation} \overline{\varv_{\rm conv}} = \frac{\overline{F_{\rm conv}}}{\left< p + P + e + E \right>} \label{eq: v_conv}. 
\end{equation} 
As discussed in detail in \citet{Debnath2024}, it is important to realize that this convective velocity associated with energy transport is not the same quantity as the turbulent velocity associated with momentum transport defined above; that is, $\overline{\varv_{\rm conv}} \ne \overline{\varv_{\rm turb}}$. As shown in Figure \ref{fig: vturb_vconv} and Table \ref{table: output}, the convective velocity can be up to two orders of magnitude lower than the turbulent one for low-luminosity models.

We can assess the relative importance of energy transport by enthalpy (often called convection) throughout the stellar atmospheres in our models by evaluating $\overline{F_{\rm conv}}/\langle F_{\rm tot}\rangle$ as a function of atmospheric depth. 
From this profile (see e.g. \citealt{Debnath2024}), it is clear that the region where such energy transport can play a non-zero role coincides with the temperature range of the iron opacity bump. 
This agrees with the results of previous massive-star simulations of this region \citep{Jiang2015, Schultz2023, Moens2022}.

\begin{figure} 
\centering 
\includegraphics[width=\hsize]{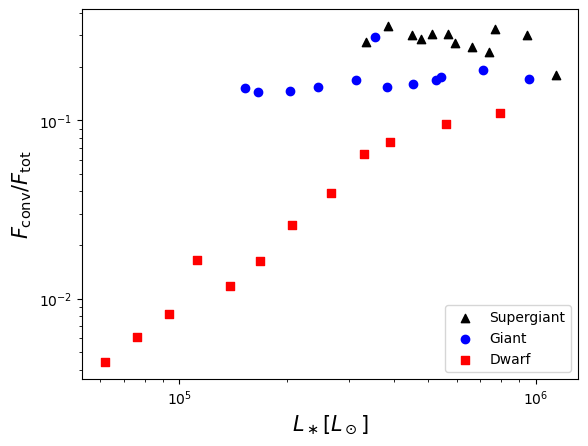} 
\caption{Maximum value of the ratio of convective to total energy flux as a function of stellar effective temperature for all of the simulations in our grid. Red squares indicate the series of dwarf models, blue circles the series of giants, and black triangles the series of supergiants.} 
\label{fig: Fconv_grid} 
\end{figure}

Additionally, we can see from Fig. \ref{fig: Fconv_grid} (see also Fig. 15 in \citealt{Debnath2024}) that such convective energy transport is not very efficient in these O star atmospheres as compared to convective regions in low temperature stars. For the dwarf models, at the peak of the iron bump, not more than $\sim 10 \%$ of the total flux is carried by enthalpy, with values on the order of $\sim 1 \%$ for many stars of the lower luminosity dwarfs (see Fig. \ref{fig: Fconv_grid}). The giants all reach peak values around $\sim 20 \%$, while the convective flux contribution in the supergiant peaks around $\sim 30 \%$. 
Since the kinetic and gravitational fluxes are also small, the dominant energy transport mode through our simulated turbulent atmospheres is radiative diffusion.

A detailed analysis \citep{VanderSijpt2025} further shows that even though our O star simulations are convectively unstable in this region, structure growth does not generally follow the expected growth behavior of unstable, radiation-modified gravity modes. 
Atmospheric motions here are rather driven by the radiation force and the opacity behavior, in contrast to classical convection where hot, energy-rich cells of gas are accelerated upward due to buoyancy. 

\subsection{Atmospheric structure} \label{ssec: inflation}

In this section, we look at how the global radial atmospheric structure is affected by multi-D effects. More precisely, the average radial density and temperature profiles, and with that the average photospheric radii and effective temperatures of the models. The 2D models presented here differ from simple, purely radiative 1D hydrostatic models in three main ways, described below  

First, purely radiative 1D models show a tendency to statically inflate, as the radiative flux over the iron bump  approaches the Eddington limit \citep{Grafener2011,Langer2015a}. As such, the effective atmospheric scale height grows to infinity and, as a result, the pressure and density profiles would plateau as a function of radius, until the opacity decreases due to a temperature drop. In the models presented in this work, part of the radiative flux gets converted to convective flux. Since this convective flux does not contribute to the radiation force, the local Eddington factor will slightly decrease. With a lower Eddington factor
the effective scale height of the atmosphere will also be lower and, as such, the density and pressure profiles will decrease more rapidly as a function of radius. 
In stellar structure models, this is often imitated by estimating the convective flux using mixing-length theory (MLT) \citep{Joyce2023, Bohm-Vitense1958}. MLT depends on a free parameter, the mixing length, $\alpha_{\rm MLT}$, which is typically calibrated from observations or simulations. 
The effect of either using MLT (or allowing convection in multi-D setup such as in the work presented here) is that the atmosphere shrinks as compared to a simpler purely 1D radiative model. 

A second important effect found in the models presented here is the presence of the turbulent velocity field along with its turbulent pressure. In a purely radiative 1D hydrostatic model, the subphotospheric atmosphere is supported by gas and radiation pressure. In a multi-D model, the turbulent pressure again increases the the atmospheric scale height. As also discussed in \citet{Debnath2024, Goldberg2022} and confirmed in the study detailed here, the turbulent pressure for O stars can be on the same order of magnitude as the radiation pressure right below the photosphere.
While the convective flux and its effect on the scale height is mostly important in the deeper atmosphere, the turbulent pressure support peaks very close to the photosphere. As a result, the average density, gas pressure, and temperature profiles of the atmospheres are a lot less steep at the photosphere than compared to the same profiles in a 1D model. The more gentle sloping density profile means that the photosphere again moves outwards as compared to a 1D model. As a result, the effective temperature will be lower (see also, \citealt{Gonzalez-Tora2025a}).

These two mechanisms, the lack of static inflation due to convection in the deeper atmosphere and the turbulent pressure support near the photosphere have opposite effects on the photospheric radius and effective temperature. Whichever effect prevails depends heavily on the opacity structure and the atmospheric dynamics in the model. Models with a higher Eddington ratio will be more prone to static inflation and models with a lower surface gravity are more effective at converting luminosity to convection. Turbulent velocity and pressure both seem to be primarily correlated with the Eddington ratio.

A third noticeable effect in the models presented here is the lack of a density inversion. It appears that such structures are unstable in a multi-D setup and, as such, they break down via a Rayleigh-Taylor like mechanism \citep{VanderSijpt2025}.

The change in photospheric radius for the full grid of models is shown in Fig. \ref{fig: Rphot_grid}. 
This effect could lead to a mismatch between the observed position of a star on the HRD and the position calculated by stellar evolution codes, such as  MESA \citep{Paxton2019} or GENEC \citep{Eggenberger2008}, which do not fully take into account these complexities.

\begin{figure}
\centering
\includegraphics[width=\hsize]{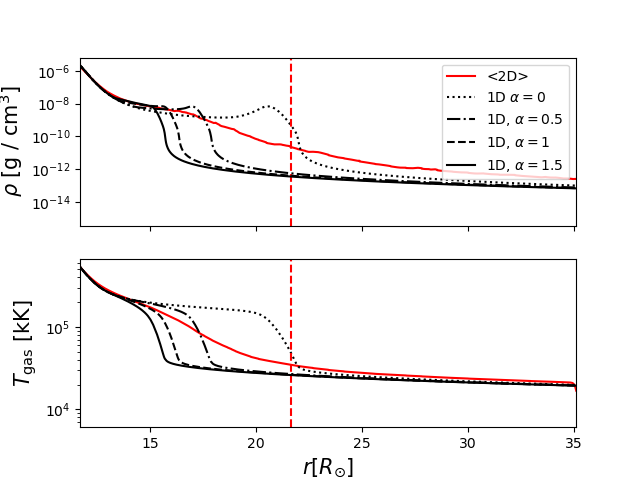}
  \caption{
  Average density (top) and temperature (bottom) profiles of the 2D simulation (red) of model G10, together with 1D 
  models with different mixing lengths that all match the density and temperatures at the bottom boundary of the numerical domain (black dotted). The red dashed line indicates the $<2D>$ photosphere.
  }
     \label{fig: 1D_prof}
\end{figure}

\begin{figure}
\centering
\includegraphics[width=\hsize]{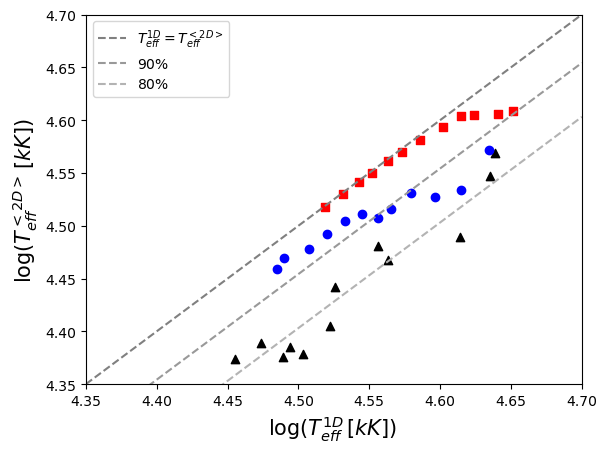}
  \caption{Change in effective temperature due to inflation for all of the simulations in our grid. Red squares indicate the series of dwarf models, blue circles the series of giants, and black triangles the series of supergiants. 
  }
     \label{fig: Teff_grid}
\end{figure}

\begin{figure}
\centering
\includegraphics[width=\hsize]{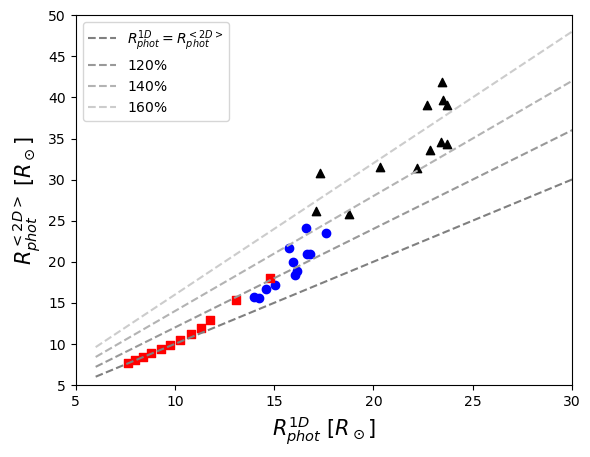}
  \caption{Comparison of the photospheric radius in a 2D model and a 1D purely radiative model with a closely matching bottom boundary. Red squares indicate the series of dwarf models, blue circles the series of giants, and black triangles the series of supergiants. 
  }
     \label{fig: Rphot_grid}
\end{figure}

Figure \ref{fig: 1D_prof} shows the average density and temperature profile of 2D model G10 as compared to four 1D hydrostatic models with different values for $\alpha_{\rm MLT}$. One can clearly see the lack of density inversion, lack of static inflation, and how the average 2D profiles are less steep near the photosphere.
In Figs. \ref{fig: Rphot_grid} and \ref{fig: Teff_grid}, we compare the relaxed effective temperatures and photospheric radii to those computed from the same 1D model where we chose $\alpha_{\rm MLT}=1.5$. 
While the mixing length parameter has been well gauged by means of multi-D RHD calculations for solar like stars \citep{Magic2013}, red supergiants \citep{Freytag2012}, and asymptotic giant branch (AGB) stars \citep{Freytag2017}, 
as of yet there is no consensus value to be used in these types of regions. 
Although \citet{Debnath2024} found that this is probably too high for one of their models, the value of 1.5 is used in previous literature (e.g. \citealt{Cantiello2009,Jiang2015}).
As compared to the 1D models with $\alpha_{\rm MLT}=1.5$ 
, we can see a clear increase in photospheric radius, and decrease in the effective temperatures of the 2D models due to the inflation by means of turbulent pressure. 

\subsection{Wind structure}
\label{ssec: Mdot}
\begin{figure}
\centering
\includegraphics[width=\hsize]{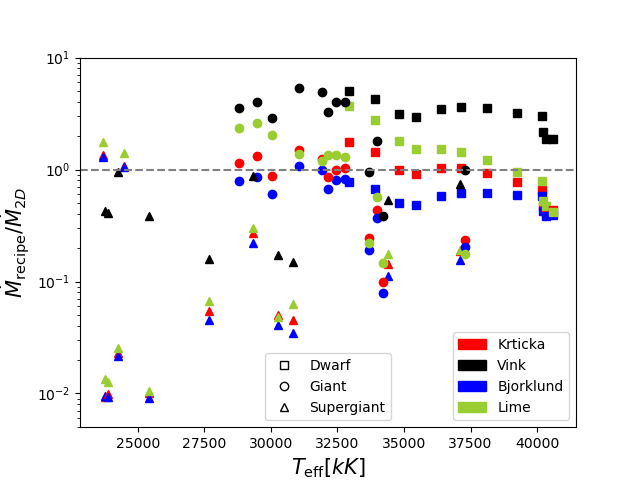}
  \caption{\ Mass-loss rate for the entire grid of models, compared to mass loss rates predicted by the \citet{Vink2001}, \citet{Bjorklund2023}, and \citet{Krticka2024} recipes, and the LIME \citep{Sundqvist2025} algorithm. Squares indicate the series of dwarf models, circles the series of giants, triangles the series of supergiants.
  }
     \label{fig: Mdot_recipe}
\end{figure}

While calculating mass-loss rates is not the focus in this paper, we can nonetheless compare the mass-loss rates from our 2D models with some frequently used predictive formulae based on 1D stationary models to gauge the global physical accuracy of our grid of models. The recipes we selected for comparison are \citet{Vink2001}, \citet{Bjorklund2023}, and \citet{Krticka2024}. 
Additionally, we carried out a comparison with LIME
\footnote{\url{https://lime.ster.kuleuven.be/}}, a zero-dimensional approach for estimating mass loss rates based on estimating the critical point \citep{Sundqvist2025}. 
A comparison between the average mass loss rate from the 2D models and the recipes is shown in Fig. \ref{fig: Mdot_recipe}. Based on the approach that the different recipes have taken, it is expected that our models are best described by \citet{Krticka2024}. 
Both our models and the \citet{Krticka2024} models are based on computing a local line-force by brute force summing over a set of lines in the Sobolev approximation, which is then used in the momentum equation. 
For most of the supergiants (triangles in Figure \ref{fig: Mdot_recipe}), the recipes underestimate the mass loss rate. This may be because the 2D models find an extra driving force in addition to radiation force in the form of a strong turbulent pressure gradient. This is prevalent mainly in the supergiants, as these models feature higher values of the turbulent pressure.

The average supersonic wind structure is quite well fit by a $\beta$-velocity law,
\begin{equation}
\varv(r) = \varv_\infty \left(1 - \frac{R_0}{r} \right)^\beta, 
\end{equation}
as can be seen in Fig. \ref{fig: Beta_vel} for model G1.  This is unsurprising as the wind is described by driving that is very similar to a modified CAK formalism, which has a $\beta$-law as an (approximate) solution.
However, it should be noted that in a more physical 2D simulation, wind material follows a broad distribution of velocities at every single radius (e.g., \citealt{Moens2022,Debnath2024}). 
The wind also hosts a strong anti-correlation between the radial velocity and gas density, where heavier material flows on average slower than low-density material, which is accelerated by line-driving more efficiently. This is also in line with a similar anti-correlation reported in \citet{Moens2022} and \cite{Debnath2024}. Another feature that deviates significantly from 1D stationary predictions can be seen in the transition region of Fig. \ref{fig: Beta_vel}, right before the onset of the average supersonic wind, where the unweighted average flow velocity often is slightly negative before rapid outwards acceleration starts. A closer inspection reveals that in these regions, the radial velocity is correlated with density 
as a result of the Rosseland opacity behavior;
as such, the outflowing material transports more mass than the infalling material and the density-weighted radial velocity is roughly constant at 0.

\section{Spectroscopic consequences}

\label{sec: Spectroscopic properties}
We go on to explore what first order effects the complex density and velocity fields of the turbulent atmosphere have on the broadening of photospheric spectral absorption lines. For this purpose, we ran a simplified RTE solver to calculate emergent intensities for several spectral lines from our 2D models and we compared those to spectral lines computed from matching hydrostatic, lateral-, and time-averaged 1D models. Appendix \ref{app: averaging} holds some more details on how the RTE solutions along multiple rays in multiple snapshots are combined into one.

\subsection{RTE Solver} \label{ssec: RTE solver}
Approximated synthetic spectral lines can be computed from the RHD models by integrating the steady state RTE along a set of purely radial rays. Here, one ray is used for each lateral cell on the finest level of the AMR grid,

\begin{equation}
    \partial_r I_\nu = \chi_\nu \left(S_\nu - I_\nu \right)
.\end{equation}

The integrated, frequency-dependent intensities can then be added together from different rays to approximate the synthetic spectral line.
This simplified, one-ray approach allows us to make relative comparisons between line calculations in a quick and easy approach. 
In our simple RTE solution, we assume an LTE source function for both the line and continuum, where the gas temperature is inferred from the ideal gas law. 
The absorption coefficient is computed as 
\begin{equation}
    \chi_\nu = q_{\rm line}(\rho,T_g) \phi_\nu(v) + \rho \kappa_e,
\end{equation}

where the integrated line strength $q_{line}$ depends on the line occupation numbers obtained from solving the Saha-Boltzmann equations \citep{Poniatowski2022}. 
The line profile function, $\phi_\nu$, has a thermal width and is shifted in frequency by the local velocity field due to Doppler shifting \citep[see, e.g., Eq.\,5 in][]{Hennicker2022}. The intensity for the continuum is computed in a similar fashion, but here the absorption coefficient is given solely by electron scattering.

We then assessed the line profiles by evaluating the emergent line intensities normalized to the nearby continuum. We calculated a set of optical lines from our underlying Munich atomic data base \citep{Pauldrach1994, Puls2005}, which are often used in spectroscopic studies of O stars \citep{Hawcroft2021, Verhamme2024, Brands2022, Brands2025, Backs2024a, Backs2024}. 
These lines are formed at different depths of the turbulent atmosphere. Therefore, different lines might trace different regions showing differences in the turbulent velocity field through broadening, shifts, or asymmetries.
Specifically for the purposes of this study, we used the optical photospheric lines from the study by \citet{Hawcroft2021}, except for the Stark-broadened Balmer lines. We also looked at \ion{C}{iv}\,5801 to include a relevant Carbon line that is not dominated by the wind, along with \ion{Si}{iii}\,4552, which is often used 
for estimates of photospheric velocity fields based on 1D models (e.g., \citealt{McErlean1998}).\footnote{These carbon and silicon lines are known to be a doublet and triplet respectively, but here they are handled as if they were singlets.}.

\subsection{Effects of photospheric velocity fields on optical absorption line profiles} \label{ssec: broadening}

\begin{figure}
\centering
\includegraphics[width=\hsize]{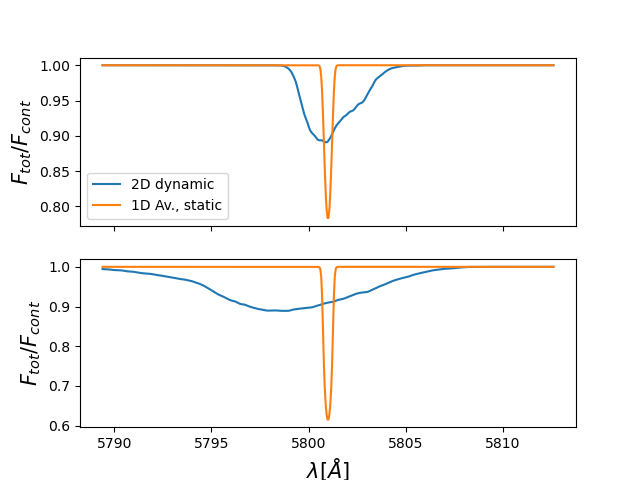}
  \caption{Synthetic CIV 5801 lines computed for the the 2D, dynamic atmosphere and wind models (blue), and the laterally averaged models without any velocity field (orange).  The result for the least turbulent giant model, G1 is in the top panel and in the bottom panel for the most turbulent supergiant, S12.   
  }
     \label{fig: 1D_to_broad_prof}
\end{figure}

The widths of the calculated absorption lines are directly related to the amplitudes of Doppler shifts introduced by the atmospheric velocity fields in our dynamic O star simulations. To assess this overall effect, we averaged the line profiles computed from the many different snapshots of our 2D models and then compared these to corresponding profiles calculated from 1D density and temperature averages and with all atmospheric velocity fields set to zero. 
As can be seen in Fig. \ref{fig: 1D_to_broad_prof} for a \ion{C}{iv}\,5801 line in models G1 and S12, the 2D dynamic line is broadened and has a higher equivalent width than the 1D-averaged static line profile. The broadening is stronger for the more turbulent supergiant model. Compared to the 1D static lines, the 2D dynamic lines is also more structured due to the complex velocity field, even despite the time averaging.
The synthetic lines found by the global 3D radiative transfer calculations done in \citet{Delbroek2025} also seem to be a lot smoother, but it is unclear wether this is due to the smoother 3D structure or a lower spatial resolution of the radiative transfer calculations.

To quantify the line broadening process, we compared the synthetic lines calculated from our 2D models, $\phi_{\rm 2D}$, to synthetic lines calculated from the 1D average density and temperature profile, $\phi_{\rm av.}$, where we completely omitted the velocity field. Then, we broadened the latter profile by convolving it with a Gaussian broadening profile, $\psi(\lambda),$ in velocity space, 

\begin{equation}
    \phi_{\rm broad.}(\lambda) = \int_{-\infty}^{\infty} \phi_{\rm av.}(\lambda) \psi_{\rm broad.}(\lambda - \lambda') d\lambda'
.\end{equation}

The relevant broadening parameter can then be determined by finding the value of $\varv_{\rm broad.}$, where  $\phi_{\rm 2D}$ fits $\phi_{\rm broad.}$ best.
The quantity $\varv_{\rm broad.}$ is 
related
to the macro-turbulent velocity, $\varv_{\rm macro}$, for an isotropic Gaussian velocity distribution  \citep[e.g.,][]{Gray2005}; however different definitions are often used in the literature.
The broadening profile here is defined as
\begin{equation}
    \psi(\lambda) \coloneqq \frac{A}{\sqrt{2\pi}} \exp{\left( \frac{-(\lambda - \lambda_0)^2}{2\sigma^2} \right)},
\end{equation}
with $\lambda_0$ denoting the line center, $A$ a normalization constant used in the fitting process and $\sigma \coloneqq \lambda_0 \varv_{\rm broad.}/c$ the width of the profile in wavelength space.

During this approach, we allow different lines to have different values of the broadening parameter. As such, we can also assess the range of values for this parameter within one stellar object.

\begin{figure}
\centering
\includegraphics[width=\hsize]{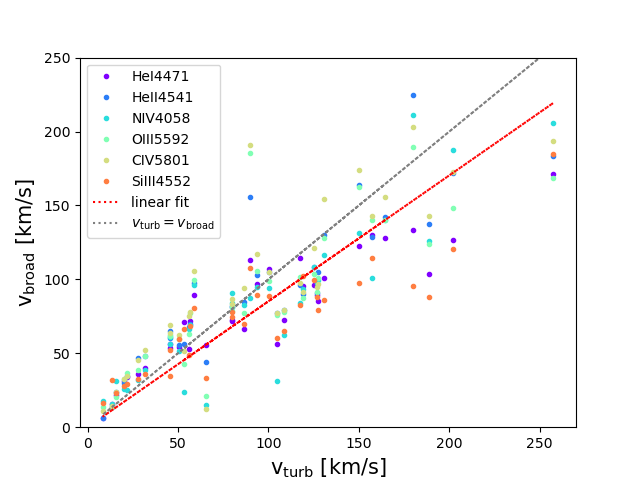}
  \caption{Correlation between the maximum, sub-photospheric turbulent velocity, and the one standard deviation width of the best fit turbulent broadening function. Different colors represent different atomic transitions and different locations on the horizontal axis correspond to different stellar models. 
  }
     \label{fig: vmacro_vturb}
\end{figure}

As shown by Fig.  \ref{fig: vmacro_vturb}, there is a strong correlation between the broadening of the synthetic line profiles, $\varv_{\rm broad}$, and the dynamic quantity, $\varv_{\rm turb}$. 
Even more so, the average scaling factor between $\varv_{\rm broad}$ and $\varv_{\rm turb}$ seems to be on the order of unity, with: $\varv_{\rm broad} \approx 0.85 \varv_{\rm turb}$.
A similar near-unity relation was also found for a similar calculation done on a single 3D model, using full 3D radiative transfer in \citet{Delbroek2025}.
Within one single model, there is a spread of about a factor of two in broadening velocity between different atomic transitions. This is due to the different regions within the atmosphere where the transition  interacts efficiently with the radiation field due to the local level populations.

The work here focuses solely on the turbulent broadening of Doppler profiles. An additional point of future research would be to investigate the effect of sub-surface turbulence on Stark broadened Voigt profiles. While this research has not been carried out here, we could expect a similar line broadening due to the turbulent velocity fields. This broadening, however, would be counteracted by the thinner profiles that would be produced by the turbulent pressure supported atmosphere, which lowers the need for gas pressure and thus Stark broadening to maintain hydrostatic equilibrium. Initial work on this has been performed by \citet{Gonzalez-Tora2025a,Gonzalez-Tora2025} by including turbulent pressure support in 1D CMF calculations. As of now, it is not yet known which of the two effects will be stronger. It is clear from their study, however, that turbulent effects can significantly alter derived spectroscopic masses. In addition to the line broadening, there are quantifiable effects on the equivalent width, shift from line center, and asymmetries in the spectral lines (discussed in Appendix \ref{app: EW}). 

\section{Discussion} \label{sec: Discussion}

The models and 
line profile results presented in this work have strong implications for the interpretation of observations of massive stars. 
We focus the discussion below on some of those implications, along with some limitations of the methods that we have used.  

\subsection{Model limitations and future improvements}

\paragraph{2D versus 3D.}\label{sssec: 3D}
While turbulence is inherently a 3D process, we opted to use 2D models in this work to reduce the computational costs, while also allowing us to still break up spherical symmetry.
While there are quantitative differences between 2D and 3D models, previous comparisons \citep[e.g.,][]{Moens2022} of similar models showed that the qualitative structure and trends remain the same. A more in-depth comparison between 2D and 3D massive-star atmosphere models for O stars is currently in preparation and will be presented in a future work.

\paragraph{FLD and opacity limitations.}
Two important limitations of the RHD method employed for the models in this paper are the assumptions of: (1) the analytic FLD closure relation, and (2) $\kappa_E = \kappa_B = \kappa_F $. Both of these problems have already been discussed in previous publications by the authors \citep{Moens2021, Moens2022, Debnath2024}. In this work, we  chose to focus on the optically thick atmosphere and photospheric region, rather than the wind. 
Especially in the optically thick atmosphere, these two concerns cited above are minor, 
as the atmosphere is close to being fully diffusive and, as such, the opacity is dominated by the optically thick Rosseland mean contribution. It is true, however, that around the photosphere and in the overlying wind, we are in need of better heating and cooling opacities. Furthermore, we ought to take extra care when  interpreting FLD-fluxes as the FLD-closure in the transition region from diffusive to free streaming is only an approximation. 
Finally, the straight sum between Rosseland mean opacities and the line-driving opacity multiplier, as prescribed in the hybrid opacity scheme \citep{Poniatowski2021} is possibly double counting some minor line opacity contributions.

\paragraph{RTE solution.}
Before comparing the synthetic absorption lines calculated in Sect. \ref{sec: Spectroscopic properties} to observations directly, there are some steps that need to be undertaken: 1) the absorption lines here are calculated for a planar atmosphere without sphericity effects, which prevented us from computing flux profiles and lines significantly affected by the wind outflow; and 
2) the population numbers in the atmosphere in O stars will be influenced by NLTE effects \citep{Auer1969}. 
While the LTE approximation likely suffices for computing frequency-integrated opacities and forces (as shown by \citealt{Poniatowski2021}), it will not suffice for frequency-dependent radiative transfer. A future improvement is to post-process RHD models such as those discussed in this paper with full 3D NLTE radiative transfer tools, as described by \citet{Hennicker2020}. 
Very recently, first results using these tools on these types of simulations were published by \citet[under review]{Delbroek2025}.
While this will be important for direct comparisons to observations, 
we do not expect that the general qualitative turbulent broadening trends (as described in Sect.\,\ref{ssec: broadening}) will be significantly affected by NLTE or sphericity effects.

\subsection{Interpretation as macro-turbulence}

High-resolution spectroscopy has given the opportunity to uncover additional broadening mechanisms in massive star spectra. Besides rotational broadening, stellar pulsations \citep{Aerts2003} and turbulence \citep{Howarth2004} are two candidates that are often discussed in this context.

While it is impossible to parameterize the entire atmospheric velocity field, the micro- and macro-turbulence formalism tries to parameterize velocities in the two limits for the sizes of turbulent structures: 
Velocities of turbulent elements on scales smaller than an optical depth are treated by adding a micro-turbulent broadening term in the width of the line profile function, $\phi_\nu$, during the RTE integration; on scales greater than the optical depth, a macro-turbulent broadening function is used to convolve the solution to the formal integral \citep{Gray2005}. In that sense, micro- and macro-turbulence are not two distinct processes, but, rather, two limits on the same stochastic process that is taking place on a continuum of length scales. 

The broadening mechanism quantified in Sect.\,\ref{ssec: broadening} holds some physical connection to the macro-turbulent broadening parameter used in spectroscopy by means of 1D atmosphere models \citep{Gray2005}. 
Looking at macro-turbulent velocities derived from \citet[their Fig.\,6]{Simon-Diaz2017}, we find a clear similarity with the turbulent velocities found in our models. Here, we call attention to  the difference in luminosity, $L_\ast$, and spectroscopic luminosity (i.e., gravity-weighted flux), expressed as $\mathcal{L}_\ast \propto L_\ast/M$. 
This match is both qualitative, showing a similar trend in $\varv_{\rm turb}(L_\ast),$ as well as somewhat quantitative; the same order of magnitudes is found in both velocities for a similar range of stellar luminosities. 
Additionally, simulations using the same framework, but including magnetic fields \citep{ud-Doula2025, Narechania2025} show how strong magnetic fields can suppress turbulent dynamics underneath the photosphere, which matches the inhibited macro-turbulent velocities in highly magnetic O stars \citep{Sundqvist2013a}.

In this paper, we have opted for a simple gaussian broadening profile to keep the analysis as simple as possible. 
Observational studies have shown that in practice a lot can be learned from the exact shape of the broadening kernel \citep{Aerts2014, Simon-Diaz2014, Serebriakova2023}. 
A deeper study on the exact shape of the broadening mechanism will again have to rely on a more accurate radiative transfer \citep{Takeda2017} and is left for future work. 

The effects and presence of non-isotropic macro-turbulent motion have recently been studied for a sample of TESS OB stars, which received follow-up from the ESO UVES/FEROS Large Programs. In line with our findings for the relative velocity field dispersion components (see Sect.\,\ref{ssec: Turbulent pressure}), \citet{Serebriakova2023} found a significant contribution coming from the tangential macroturbulent broadening component in B stars. Despite large error bars, this tangential component seems to be stronger than the radial one for most of the objects in the sample. 
While we see this in our O star models, multi-D modeling efforts for B-stars are still a work in progress.

\subsection{Implications for 1D stellar structure and atmosphere codes}
As discussed in Sect. \ref{ssec: inflation} and also \citet{Debnath2024}, the addition of turbulent pressure in the stellar atmosphere affects the local total scale height and, consequently, the stellar structure near the outer layers. These are effects that are typically not taken into account, neither in 1D stellar evolution codes, nor in 1D atmosphere and wind codes. 

\paragraph{Stellar Masses.}

A significant turbulent pressure support might alter stellar masses derived from both spectroscopy and evolutionary models. Firstly, spectroscopic masses are derived by fitting for the surface gravity,  
assuming hydrostatic equilibrium from the supporting gas and radiation pressure gradients in the photosphere. If there is an additional turbulent pressure support, this pressure balance changes.
This effect has likely been observed in the galaxy for OB stars by \citet{Herrero1992}.
A more modern, theoretical example of this can be seen in the 
recent PoWR computations that do take into account such turbulent support \citep[see also discussions in \citealt{Markova2018}]{Gonzalez-Tora2025a, Gonzalez-Tora2025}. Secondly, the evolutionary mass is computed by fitting stellar evolutionary tracks through observed positions on the HRD \citep[e.g.,][]{Schneider2014}. Our 
multi-D models hint at a different connection between stellar core properties and atmospheric observables than computed from previous 1D models, even on the main sequence. The main effects that lead to this are the change of the photospheric radii and effective temperatures, as discussed in Sect.\,\ref{ssec: inflation}. To circumvent this mismatch, we are in need of a better implementation of stellar atmospheres in stellar structure and evolution codes. 
One possible way forward is to parametrize the turbulent pressure support and the MLT mixing length informed by simulations such as those presented here. 
Another possibility is the alternative formulation by \citet{Schultz2020}. However, this formalism again ignores the very important turbulent pressure support.

\paragraph{Wind structure.}
Finally, the models presented in this paper vear away from the classical two-component clumped wind picture, where the wind is assumed to be a smooth low density interclump medium with a distribution of discrete clumps on top of that.
The processes over the iron bump lead to dynamical and physically different structures as compared to the line-deshadowing instability (LDI, \citealt{OwockiRybicki1984})
, an instability prevalent in optically thin winds thought to be the origin of wind clumping. In our models, density and velocity structures seem to be more smooth and continuous as opposed to the clearly defined, strong local clumps seen in 1D LDI simulations \citep{driessenTheoreticalWindClumping2019}. 

It seems that corrections for this type of structure formation that have been implemented in 1D codes using different types of clumping mechanisms \citep{Hillier1991, Puls2006, SundqvistPuls2018} cannot  be reconciled with the broad distributions here, which clearly do not resemble two-component-like scenarios. It is possible that real physical stellar winds are affected by a superposition of both these processes, which cannot be covered by our current methods. This newly uncovered structure formation mechanism could have significant effects on the way that clumped winds are treated at present in comoving frame-based codes, namely,  predominantly based on 1D LDI-like structures. 
Finding  a solution for this in the future is not a trivial task. 

\section{Summary} \label{sec: Conclusions}

In this work, we present a grid of unified 2D O star atmosphere and wind models based upon the 
approach developed and outlined in \citet{Poniatowski2022, Moens2022, Debnath2024}. Due to instabilities in the stellar envelope, these models show structure formation in the atmospheres and winds of O stars without any a priori assumptions.
We quantified the turbulent and convective properties over the grid of models, focusing primarily on the significant role turbulent pressure plays for the support of the O star atmospheres and on the inefficiency of energy transport by means of enthalpy. We find that the characteristic atmospheric turbulent velocity in the models scale quadratically with the classical Eddington parameter.
These turbulent velocities result in turbulent pressures significantly affecting  the global atmospheric structure and scale heights. As such, both the average effective temperatures and photospheric radii are affected by multi-D effects. These effects increase with the classical Eddington parameter and lower local gravity. 

We further investigated an initial, simple proxy for quantifying non-rotational spectral line broadening effects and found a strong relation between the broadening of spectral lines in velocity space, as well as the turbulent velocities near the photosphere. This suggests that models such as the ones presented in this work can be used to predict broadening parameters for observed spectra, instead of retrieving them from fitting procedures with a high number of input parameters. 

Subsequent works in this area will be focused on the extension to 3D models and a more detailed spectroscopic analysis \citep{Delbroek2025}, as well as on further improvements of the employed RHD techniques, extensions towards lower metallicities, and accounting for the effects of stellar rotation.

\begin{acknowledgements}
    The computational resources used for this work were provided by Vlaams Supercomputer Centrum (VSC) funded by the Research Foundation-Flanders (FWO) and the Flemish Government.
    NM, JS, DD, CVdS, FB acknowledge the support of the European Research Council (ERC) Horizon Europe grant under grant agreement number 101044048, of the Belgian Research Foundation Flanders (FWO) Odysseus program under grant number G0H9218N, of FWO grant G077822N, and of KU Leuven C1 grant BRAVE C16/23/009. 
    AACS acknowledges support by the Deutsche Forschungsgemeinschaft (DFG, German Research Foundation) in the form of an Emmy Noether Research Group -- Project-ID 445674056 (SA4064/1-1, PI Sander) -- and a Research Grant under Project-ID 496854903 (SA4064/2-1, PI Sander). This project was co-funded by the European Union (Project 101183150 - OCEANS). This work was performed in part at Aspen Center for Physics, which is supported by National Science Foundation grant PHY-2210452.
    The authors would like to express their gratitude to C. Aerts for her helpful comments on the manuscript.
    The authors further like to thank the referee for their useful and constructive comments on the manuscript.
    Data analysis and visualization in this project relied on Visit \citep{Visit}, Numpy \citep{Numpy}, Scipy \citep{Scipy}, and Matplotlib \citep{Matplotlib}

\end{acknowledgements}

%
%

\bibliographystyle{aa}
\bibliography{My_library.bib}

\newpage

\begin{appendix}

\onecolumn

\section{Grid Input and output parameters} \label{app: Grid io}

In this appendix, we summarize the important input and output values for a select number of physical quantities. Table \ref{table: input} displays the main stellar parameters. The first column shows the model name used in the paper, 
the second column shows the stellar mass of the model. Columns three and four show the lower boundary density and lower boundary radius. Together with an input luminosity, this mass, density, and radius define the free parameters of the RHD atmosphere model.
The next four columns, underneath the joint header "input," show the parameters that belong to the 1D purely radiative model that was used as initial conditions: Photospheric radius, effective temperature, model luminosity and Eddington parameter. The final four columns, underneath the joint header 'output', display the these stellar parameters when recalculated from the relaxed multi-D RHD model. The photospheric radii and effective temperatures change, as explained in Sect. \ref{ssec: Average Struc}, and the luminosity and Eddington ratio are recomputed from the FLD fluxes.

\FloatBarrier

\begin{table*} [h!]
\caption{Grid member input parameters in order: model name, model stellar mass, lower boundary density and radius. 4 input parameters for the purely radiative initial conditions: photospheric radius, effective temperature, stellar luminosity and classical Eddington ratio, and, finally, the same four parameters recalculated from the multi-D RHD model.} 
\label{table: input} 
\centering 
\begin{tabular}{ll|ll|llll|llll}     
\hline\hline 
\multicolumn{2}{l}{Model}       & \multicolumn{2}{l}{Lower bound} & \multicolumn{4}{l}{Input}                        & \multicolumn{4}{l}{Output}                       \\
Name & $M_\ast$ [$M_\odot$] & $\rho_0 \, \rm [g / cm^3]$         & $R_0 \, [R_\odot]$        & $R_{\rm ph} \, [R_\odot]$ & $T_{\rm eff} \, \rm [kK]$ & $L_\ast \, [L_\odot]$ & $\Gamma_e$ & $R_{\rm ph} \, [R_\odot]$ & $T_{\rm eff} \, \rm [kK]$ & $L_\ast \, [L_\odot]$ & $\Gamma_e$ \\
\hline 
D1 & 16.6 & 7.65e-06 & 6.69 & 7.4  & 30.9 & 4.4e+04 & 0.07 & 7.6  & 32.9 & 6.2e+04 & 0.10 \\ 
D2 & 18.1 & 6.94e-06 & 7.00 & 7.7  & 31.9 & 5.6e+04 & 0.08 & 8.0  & 33.9 & 7.6e+04 & 0.11 \\ 
D3 & 20.0 & 6.38e-06 & 7.35 & 8.1  & 32.9 & 6.9e+04 & 0.09 & 8.4  & 34.8 & 9.4e+04 & 0.12 \\ 
D4 & 22.0 & 5.89e-06 & 7.72 & 8.5  & 33.7 & 8.4e+04 & 0.10 & 8.9  & 35.5 & 1.1e+05 & 0.13 \\ 
D5 & 24.3 & 5.48e-06 & 8.11 & 8.9  & 34.6 & 1.0e+05 & 0.11 & 9.4  & 36.4 & 1.4e+05 & 0.15 \\ 
D6 & 26.6 & 5.12e-06 & 8.50 & 9.4  & 35.3 & 1.2e+05 & 0.12 & 9.9  & 37.2 & 1.7e+05 & 0.17 \\ 
D7 & 29.1 & 4.52e-06 & 8.85 & 9.8  & 36.7 & 1.6e+05 & 0.14 & 10.5  & 38.1 & 2.1e+05 & 0.19 \\ 
D8 & 31.8 & 3.78e-06 & 9.17 & 10.2  & 38.5 & 2.1e+05 & 0.17 & 11.2  & 39.2 & 2.7e+05 & 0.22 \\ 
D9 & 34.2 & 3.20e-06 & 9.41 & 10.6  & 40.1 & 2.6e+05 & 0.20 & 11.9  & 40.2 & 3.3e+05 & 0.25 \\ 
D10 & 37.3 & 2.77e-06 & 9.71 & 11.1  & 41.5 & 3.3e+05 & 0.23 & 12.9  & 40.2 & 3.9e+05 & 0.27 \\ 
D11 & 46.0 & 2.22e-06 & 10.53 & 12.3  & 43.6 & 4.9e+05 & 0.28 & 15.4  & 40.3 & 5.6e+05 & 0.32 \\ 
D12 & 58.1 & 1.97e-06 & 11.71 & 13.8  & 44.8 & 6.9e+05 & 0.31 & 18.0  & 40.6 & 7.9e+05 & 0.36 \\ 
G1 & 21.1 & 3.78e-06 & 10.86 & 13.4  & 30.4 & 1.4e+05 & 0.17 & 15.7  & 28.8 & 1.5e+05 & 0.19 \\ 
G2 & 23.2 & 3.77e-06 & 11.27 & 13.7  & 30.8 & 1.5e+05 & 0.17 & 15.6  & 29.5 & 1.7e+05 & 0.19 \\ 
G3 & 24.9 & 3.38e-06 & 11.36 & 13.9  & 32.0 & 1.8e+05 & 0.19 & 16.7  & 30.0 & 2.0e+05 & 0.21 \\ 
G4 & 27.0 & 3.20e-06 & 11.58 & 14.1  & 32.8 & 2.1e+05 & 0.20 & 17.1  & 31.0 & 2.4e+05 & 0.24 \\ 
G5 & 29.2 & 3.04e-06 & 11.81 & 14.3  & 33.6 & 2.3e+05 & 0.21 & 18.3  & 31.9 & 3.1e+05 & 0.28 \\ 
G6 & 31.3 & 2.76e-06 & 11.86 & 14.5  & 34.8 & 2.8e+05 & 0.23 & 18.9  & 32.4 & 3.5e+05 & 0.30 \\ 
G7 & 33.8 & 2.53e-06 & 11.95 & 14.7  & 35.9 & 3.2e+05 & 0.25 & 20.0  & 32.1 & 3.8e+05 & 0.30 \\ 
G8 & 36.5 & 2.42e-06 & 12.15 & 15.0  & 36.7 & 3.6e+05 & 0.26 & 20.9  & 32.8 & 4.5e+05 & 0.32 \\ 
G9 & 39.1 & 2.13e-06 & 11.98 & 15.1  & 38.1 & 4.3e+05 & 0.29 & 21.0  & 34.0 & 5.2e+05 & 0.35 \\ 
G10 & 41.6 & 1.90e-06 & 11.71 & 15.3  & 39.5 & 5.1e+05 & 0.32 & 21.6  & 33.7 & 5.4e+05 & 0.34 \\ 
G11 & 49.1 & 1.64e-06 & 11.60 & 15.8  & 41.7 & 6.8e+05 & 0.36 & 24.1  & 34.2 & 7.1e+05 & 0.38 \\ 
G12 & 59.0 & 1.53e-06 & 11.99 & 16.6  & 43.2 & 8.6e+05 & 0.38 & 23.5  & 37.3 & 9.6e+05 & 0.43 \\ 
S1 & 30.2 & 2.32e-06 & 15.78 & 23.1  & 28.4 & 3.1e+05 & 0.27 & 34.4  & 23.7 & 3.3e+05 & 0.29 \\ 
S2 & 31.6 & 2.13e-06 & 15.17 & 22.6  & 29.6 & 3.5e+05 & 0.29 & 34.6  & 24.5 & 3.9e+05 & 0.32 \\ 
S3 & 34.3 & 2.06e-06 & 15.07 & 22.2  & 30.7 & 3.9e+05 & 0.30 & 39.6  & 23.8 & 4.5e+05 & 0.34 \\ 
S4 & 37.0 & 2.14e-06 & 15.65 & 22.0  & 31.2 & 4.1e+05 & 0.29 & 39.1  & 24.3 & 4.8e+05 & 0.34 \\ 
S5 & 39.3 & 2.06e-06 & 15.47 & 21.7  & 32.2 & 4.5e+05 & 0.30 & 41.8  & 23.9 & 5.1e+05 & 0.34 \\ 
S6 & 41.0 & 1.90e-06 & 14.78 & 21.1  & 33.4 & 5.0e+05 & 0.32 & 39.0  & 25.4 & 5.7e+05 & 0.36 \\ 
S7 & 43.0 & 1.76e-06 & 14.12 & 20.7  & 34.8 & 5.6e+05 & 0.34 & 33.6  & 27.7 & 5.9e+05 & 0.36 \\ 
S8 & 45.5 & 1.70e-06 & 13.86 & 20.3  & 35.8 & 6.1e+05 & 0.35 & 31.6  & 29.3 & 6.6e+05 & 0.38 \\ 
S9 & 47.9 & 1.58e-06 & 13.12 & 19.9  & 37.2 & 6.8e+05 & 0.37 & 31.4  & 30.3 & 7.4e+05 & 0.40 \\ 
S10 & 51.4 & 1.47e-06 & 12.39 & 19.5  & 38.8 & 7.7e+05 & 0.39 & 30.8  & 30.8 & 7.7e+05 & 0.39 \\ 
S11 & 58.3 & 1.42e-06 & 12.26 & 18.9  & 40.8 & 8.9e+05 & 0.40 & 27.0  & 34.4 & 9.2e+05 & 0.41 \\ 
S12 & 66.8 & 1.41e-06 & 12.65 & 18.5  & 42.8 & 1.0e+06 & 0.40 & 25.9  & 37.1 & 1.1e+06 & 0.44 \\ 
\hline 
\end{tabular} 
\end{table*}

Table \ref{table: output} lists the more complex output parameters used and discussed in the paper: The mass loss rate, maximum value of the turbulent velocity below the photosphere, the width of the broadening kernel in velocity space averaged over all lines, the convective velocity, and the relative contribution of the convective flux.

\FloatBarrier

\begin{table*} [h!]
\caption{Grid members output parameters in order: model name, mass loss rate, maximum value of the photospheric turbulent velocity, turbulent broadening parameter, convective velocity, and convective contribution to the total energy flux below the photosphere. 
The single value for $v_{\rm broad}$ was obtained by taking the median of the different broadening values for the different lines.} 
\label{table: output} 
\centering 
\begin{tabular}{r | l l l l l }     
\hline\hline 
Star name & $\dot{M} \, [M_\odot/ \rm yr]$ & $\varv_{\rm turb}^{max} \, \rm [km/s]$ & $\varv_{\rm broad} \, \rm [km/s]$ & $\varv_{\rm conv} \, \rm [km/s]$ & $F_{\rm conv}/F_{\rm tot}$ \\ 
\hline 
D1 & 1.6e-08 & 8.3 & 12.0 & 0.3 & 0.00 \\ 
D2 & 2.9e-08 & 13.3 & 14.8 & 0.5 & 0.01 \\ 
D3 & 5.7e-08 & 16.0 & 23.3 & 0.8 & 0.01 \\ 
D4 & 8.3e-08 & 21.6 & 34.3 & 1.7 & 0.02 \\ 
D5 & 1.0e-07 & 20.4 & 30.7 & 1.4 & 0.01 \\ 
D6 & 1.4e-07 & 28.1 & 37.4 & 2.3 & 0.02 \\ 
D7 & 2.1e-07 & 32.0 & 44.1 & 4.6 & 0.03 \\ 
D8 & 3.7e-07 & 45.7 & 61.2 & 8.7 & 0.04 \\ 
D9 & 5.8e-07 & 56.4 & 64.8 & 16.8 & 0.06 \\ 
D10 & 1.1e-06 & 86.7 & 79.8 & 24.9 & 0.08 \\ 
D11 & 2.1e-06 & 124.9 & 106.1 & 36.8 & 0.09 \\ 
D12 & 3.3e-06 & 164.4 & 141.1 & 80.0 & 0.11 \\ 
G1 & 7.5e-08 & 50.6 & 57.6 & 11.6 & 0.15 \\ 
G2 & 7.8e-08 & 45.7 & 58.4 & 9.8 & 0.15 \\ 
G3 & 1.7e-07 & 59.1 & 96.9 & 14.4 & 0.15 \\ 
G4 & 1.4e-07 & 56.4 & 71.0 & 14.1 & 0.15 \\ 
G5 & 2.5e-07 & 80.1 & 81.4 & 19.3 & 0.17 \\ 
G6 & 3.9e-07 & 79.8 & 80.2 & 22.8 & 0.29 \\ 
G7 & 5.0e-07 & 118.9 & 92.4 & 23.8 & 0.15 \\ 
G8 & 5.5e-07 & 93.6 & 100.1 & 31.6 & 0.16 \\ 
G9 & 1.7e-06 & 100.2 & 102.0 & 33.2 & 0.17 \\ 
G10 & 3.1e-06 & 117.6 & 96.9 & 42.3 & 0.17 \\ 
G11 & 1.2e-05 & 157.0 & 129.5 & 129.2 & 0.19 \\ 
G12 & 8.3e-06 & 188.7 & 124.9 & 175.1 & 0.17 \\ 
S1 & 1.2e-07 & 53.5 & 53.8 & 11.6 & 0.27 \\ 
S2 & 2.1e-07 & 65.4 & 27.0 & 16.6 & 0.34 \\ 
S3 & 2.8e-05 & 105.0 & 68.0 & 29.8 & 0.30 \\ 
S4 & 1.3e-05 & 89.6 & 134.3 & 19.3 & 0.29 \\ 
S5 & 3.3e-05 & 108.8 & 75.3 & 71.5 & 0.31 \\ 
S6 & 4.5e-05 & 127.3 & 98.2 & 89.4 & 0.31 \\ 
S7 & 1.1e-05 & 130.8 & 122.3 & 48.2 & 0.27 \\ 
S8 & 3.0e-06 & 126.9 & 91.2 & 54.5 & 0.26 \\ 
S9 & 2.1e-05 & 149.8 & 146.9 & 77.5 & 0.24 \\ 
S10 & 2.5e-05 & 180.0 & 196.3 & 111.8 & 0.32 \\ 
S11 & 1.2e-05 & 257.6 & 184.2 & 90.0 & 0.69 \\ 
S12 & 1.4e-05 & 201.9 & 160.1 & 238.4 & 0.18 \\ 
\hline 
\end{tabular} 
\end{table*}

\FloatBarrier

\twocolumn

\section{more spectroscopic consequences}

\FloatBarrier

\subsection{Time and spatial averaging for spectral synthesis} \label{app: averaging}

\begin{figure}
\centering
\includegraphics[width=\hsize]{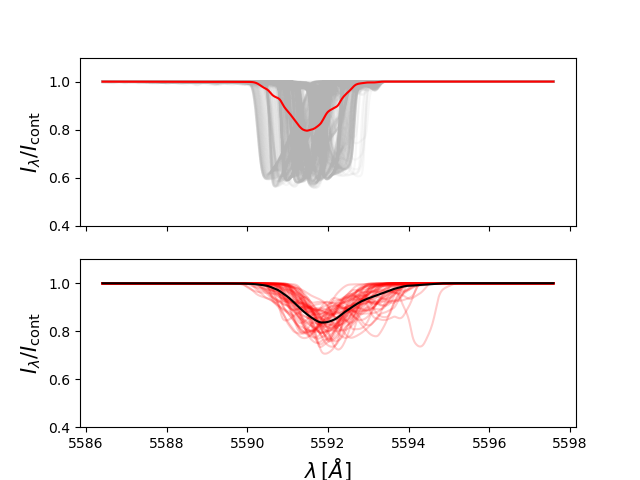}
  \caption{Temporal and lateral averaging of the radiative transfer routine. The top panel shows contributions from the 512 separate radial rays from a single snapshot (grey), along with their average (red). The bottom panel shows the 60 averages of 60 snapshots (red), together with the average of averages (black). The line corresponds to the OIII5592 transition in dwarf model D6.}
     \label{fig: Line_prof_sum}
\end{figure}

Strictly speaking, the procedure discussed in \ref{ssec: RTE solver} only provides emergent line profiles that are spatially resolved and viewed from above disk-center ($\mu = 1$). To obtain emergent flux profiles, we would need to include different viewing angles (in $\mu$ and $\phi$) and then integrate resulting emergent intensities over the stellar disk. Moreover, as our simulations are local,
atmospheric structures which would otherwise be stochastically softened, are over-represented in RTE-integrations of single snapshots. These effects can be seen in Fig. \ref{fig: Line_prof_sum}, where it is shown how different plane-parallel rays in a single snapshot return differently shaped line-profile contributions (top panel). In addition, multiple snapshots still return (slightly less so) different line-profile contributions. The question how much of this inter-snapshot variability is actual physical time-dependence in line-profiles, and how much of the variability would be dampened by sufficient sampling of the full geometric disk is still an open question. 
The question is even more complicated by the fact that the simulations performed in this paper are 2D simulations and not 3D. 

In \citet{Delbroek2025}, we investigated these questions by populating a full spherical atmosphere with local 3D Cartesian box simulations. As such the radiative transfer equation can be solved accounting for full sphericity effects.
This will help us figure out whether or not we can resolve time-dependent line-profile variation such as those observed for massive O stars \citep{Aerts2003,Aerts2014,Markova2005,Martins2015,Simon-Diaz2024}. 

\subsection{Change in EW, line shift, and asymmetries} \label{app: EW}

In addition to normalized broadening, we also find a change in the equivalent width and a shift from line center of the synthetic lines when comparing simplified 1D to our 2D models. A change in equivalent width is not covered by the macro-turbulence mechanism and requires either the inclusion of micro-turbulence or the alteration of elemental abundances. The latter was the case in the sun which led to the renewed \citet{Asplund2009} solar abundances. 

\begin{figure}
\centering
\includegraphics[width=\hsize]{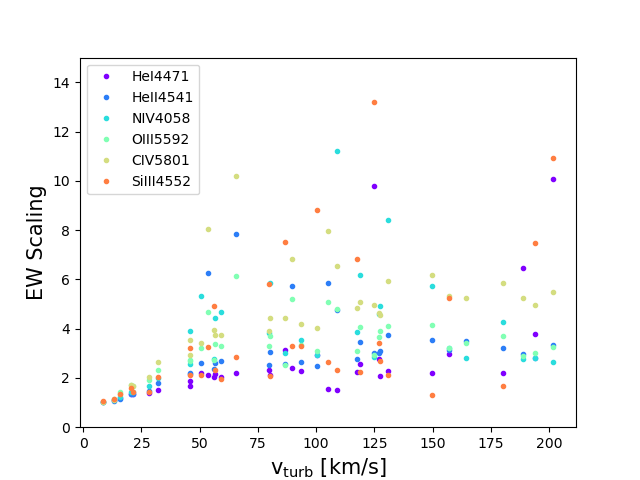}
  \caption{Increase in EW when going from a line computed from a 1D hydrostatic model, to a structured 2D model. Different colors represent different atomic transitions, and different locations on the horizontal axis correspond to different stellar models.}
     \label{fig: EW}
\end{figure}

In the models presented here, most of the equivalent widths increase with a factor on the order of $\sim 3$ between the static 1D average and the 2D dynamic atmospheres, with higher factors of the EW-changes around $\sim 10$ for some lines in the more turbulent models. Figure \ref{fig: EW} shows the change in EW for all transitions in all models as a function of their maximum sub-photospheric turbulent velocity.
Although less clear than for the turbulent line broadening seen in Fig. \ref{fig: vmacro_vturb}, there is some correlation between the change in EW and the turbulent velocity properties. Here, however, there is a quite significant spread between different atomic transitions.

For lines far from saturation, this change in EW will directly scale with the elemental abundances. In 1D model atmospheres, this is often mimicked by introducing micro-turbulence.

The asymmetry and shift in line center seen in both panels of Fig. \ref{fig: 1D_to_broad_prof} can be attributed to the distribution of radial velocities over the stellar surface near the photosphere. 

The line shift and asymmetries correspond to the velocity distribution of the gas in the line forming region near the photosphere. An on average redward shift indicates that in terms of surface area more material is infalling rather than outflowing. Due to the compressibillity of the gas this does not have to denote a net inflow of gas. A blueward shift of the line denotes a surface area that is mostly outflowing rather than infalling. Over the grid, it appears that high turbulence models are more prone to an average blueshift, while low turbulence models feature more redshifted lines.
This is confirmed by the volume-weighted velocities near the photospheres in the RHD models. See e.g. Fig. \ref{fig: Beta_vel}, which shows the infall for a low turbulence model. High turbulence models on average also have a stronger wind and thus more outflow.
The structured nature of the flow in the line-formation region is responsible for the asymmetries in the profile.

\FloatBarrier

\newpage

\section{Timescales} \label{app: timescales}

The simulation presented in this paper look at variable wind and atmosphere dynamics on a dynamical timescale. However, for the atmosphere solutions to relax to a stochastically steady-state structure, the deep layers of the simulation relax on a thermal timescale. In this section, we give a brief overview of how the typical values of the dynamical and thermal timescales vary for our grid of models. For the dynamical timescale, we can distinguish between typical dynamical velocities in the turbulent atmosphere and the mostly radial wind. Inspired by \citet{Debnath2024}, the dynamical timescales are estimated  (see also \citealt{Freytag2012}) via
\begin{equation}
    \tau_{\rm d,a} = \frac{H^{\rm tot}_{\rm p}}{\varv_{\rm turb}}
,\end{equation}

for the atmosphere, where the maximum turbulent velocity below the photosphere is used instead of the sound speed as a characteristic velocity, and the total pressure scaleheight of the average structure is used as a characteristic length scale. Here, the total pressure scaleheight $H^{\rm tot}_p = H^{\rm gas}_p p_{\rm tot}/p_{\rm gas}$ is the gas-pressure scaleheight weighted with the total pressure over gas pressure. The gas pressure scaleheight is defined as $H^{\rm gas}_p = c_{\rm iso}^2/g$, with $c_{\rm iso}$ the local isothermal soundspeed,

\begin{equation}
    \tau_{\rm d,w} = \frac{R_{\rm ph}}{\varv_\infty}    
,\end{equation}

is used for the wind dynamical timescale, where the best fit terminal velocity is used as a characteristic velocity and the photospheric radius as a characteristic length.

The atmospheric dynamical timescale is on the order of $1 \, \rm hrs$ to $10 \, \rm hrs$, with the dwarf models on the lower end. The wind dynamical timescale is typically a little lower than that and varies from $\sim 0.5 \, \rm hrs$ to $\sim 5 \, \rm hrs$ over the different models.

For estimating the thermal timescale of the simulated part of the optically thick atmosphere, \citet{Grassitelli2016}, we follow

\begin{equation}
    \tau_{\rm th} = \frac{G M_\ast M_{\rm conv}}{R_{\rm ph} L_\ast},
\end{equation}

where $M_{conv}$ is the mass contained in the turbulent region of the simulation. 
This timescale estimates the time it takes for the gravitational energy stored in the atmosphere to be radiated away in stellar luminosity. $M_{conv}$ is used instead of the complete envelope mass, as the radiative zone of the models is already relaxed from the model initial conditions onwards. The thermal timescales for each model are between one and two orders of magnitude larger than the atmospheric dynamical timescale. They reach from $\sim 10 \, \rm hrs$ for the dwarf models to $\sim 500 \, \rm hrs$ for some of the giants. First relaxing the simulation over a couple of thermal timescales ensures that the atmospheric structure has reached a thermally relaxed state. In the work presented here, we opt for relaxing over approximately $10 \, \tau_{\rm th}$. Next, to sample the variable atmosphere trough time, snapshots have been taken every equidistantly in time, once $0.1 \, \tau_{d,a}$. experience has shown that $\sim 60$ snapshots suffice to cover clear and smooth average profiles for the important studied quantities such as density, velocity, and turbulence. Additionally, over this time frame, the variable wind mass loss rate is well sampled.

\end{appendix}

\end{document}